\input epsf.sty
\documentclass[12pt]{article}

\begin{document}        
\pagestyle{empty}
\renewcommand{\thefootnote}{\fnsymbol{footnote}}

\begin{flushright}
{\small
SLAC--PUB--8162\\
May 1999\\}
\end{flushright}
 
\vspace{.8cm}

\begin{center}
{\bf\large   
Measurements of $Z^0$ to Heavy-quark couplings 
at SLD\footnote{Work supported by
Department of Energy contract  DE--AC03--76SF00515 (SLAC).}}

\vspace{1cm}
{\bf
Masako Iwasaki
} \\
\vskip 0.2cm
{\it
Department of Physics, University of Oregon, Eugene,
OR 97403
} \\
\vskip 0.6cm
{\bf
Representing the SLD Collaboration$^{**}$
} \\
\vskip 0.2cm
{\it
Stanford Linear Accelerator Center, Stanford University,
Stanford, CA  94309
}
\medskip
\end{center}
 
\vfill

\begin{center}
{\bf\large   
Abstract }
\end{center}
We present measurements of $Z^0$ to heavy-quark coupling electroweak 
parameters, $R_b$, $R_c$, and parity-violation parameter $A_c$, from SLD.
The measurements are based on approximately 550k hadronic $Z^0$ events
collected in 1993-98.
Obtained preliminary results of $R_b$ and $R_c$ measurements are
$R_b = 0.2159 \pm 0.0014 \pm 0.0014$ and 
$R_c = 0.1685 \pm 0.0047 \pm 0.0043$.
In the $A_c$ measurement, we use four methods to determine the
initial-quark charge:
combined Kaon charge and Vertex charge, lepton,  
exclusively reconstructed D*, D-mesons, and a new method using 
inclusive soft-pion from D*.
The preliminary results of these four methods were combined to give
$A_c = 0.634 \pm 0.027$.

\vfill

\begin{center} 
{\it Presented at the American Physical Society (APS) Meeting of 
the Division of Particles and Fields (DPF 99), 5-9 January 1999, 
University of California, Los Angeles} 
\end{center}

\newpage

 
 
%
\pagestyle{plain}

\section{Introduction}               

In the Standard Model, the electroweak interaction has both
vector ($v$) and axial-vector ($a$) couplings. 
Measurements of two independent parameters, 
the ratio of widths, $R_f$, and the parity-violation parameter, $A_f$, at the 
$Z^0$ resonance probe combinations of these two couplings 
of the $Z^0$ to fermions,
\begin{eqnarray}
R_f &=& \frac{\Gamma(Z^0\rightarrow f\bar{f})}
             {\Gamma(Z^0\rightarrow Hadrons)} 
  = \frac{{v_f}^2+{a_f}^2}{\sum_{i}^{udscb} ({v_i}^2+{a_i}^2) }
    \nonumber\\
A_f &=& \frac{2{v_f}{a_f}}{{v_f}^2-{a_f}^2}\ . \nonumber
\end{eqnarray}
The parameter $R_f$ measures $Zf\bar{f}$-coupling strength compared to other 
quark flavors, while $A_f$ expresses the extent of parity violation at
the $Zf\bar{f}$ vertex.
These measurements provide sensitive tests of the Standard Model.
                                           
The measurements described here are based on a 550k $Z^0$-decay data
sample taken in 1993-98 at the Stanford Linear Collider (SLC),
with the SLC Large Detector (SLD). 
A general description of the SLD can be found elsewhere\cite{sld}.
Polarized electron beams, a small and stable SLC interaction region, and 
the excellent CCD vertex detector\cite{vxd3} provide precision
electroweak measurements, especially in the heavy-quark sector.

\section{Flavor Tagging}
Topologically reconstructed mass of the secondary vertex\cite{masstag}
is used by many analyses at the SLD for heavy-quark tagging.
To reconstruct the secondary vertices, the space points 
where track density functions overlap are searched in the 
3-dimensional space.
Only the vertices that are significantly displaced from the primary
vertex (PV) are considered to be possible B- or D-hadron
decay vertices.
The mass of the secondary vertex is calculated using the tracks that
are associated with the vertex. 
Since the heavy-hadron decays are frequently accompanied by neutral particles,
the reconstructed mass is corrected to account for this fact.
By using kinematic information from the vertex flight path and
the momentum sum of the tracks associated with the secondary vertex,
we calculate the $P_T$-corrected mass $M_{P_T}$ 
by adding a minimum amount of missing momentum to the invariant mass. 
This is done by assuming the true momentum of heavy hadron is 
in a direction which minimizes the amount of transverse 
momentum added to the momentum sum of the tracks associated 
with the secondary vertex, and given by
$$M_{P_T} = \sqrt{{M^2}_{VTX} + {P_T}^2} + |P_T|,$$
where $M_{VTX}$ is the momentum sum for the tracks associated with the
reconstructed secondary vertex.
In this correction, vertexing resolution as well as the PV resolution
are crucial. 
Due to the small and stable interaction point at the SLC and 
the excellent vertexing resolution from the SLD CCD Vertex detector, 
this technique has so far only been successfully applied at the SLD. 
FIG.~1-a) shows the $P_T$-corrected mass distributions for the 
data and Monte-Carlo predictions.
To select the $Z^0\rightarrow b\bar{b}$ events, 
we apply the cut of $M_{P_T}>2\ GeV/c^2$, which provides 98\% purity.
\begin{figure}[t]	
\centerline{\epsfysize 2.8 truein \epsfbox{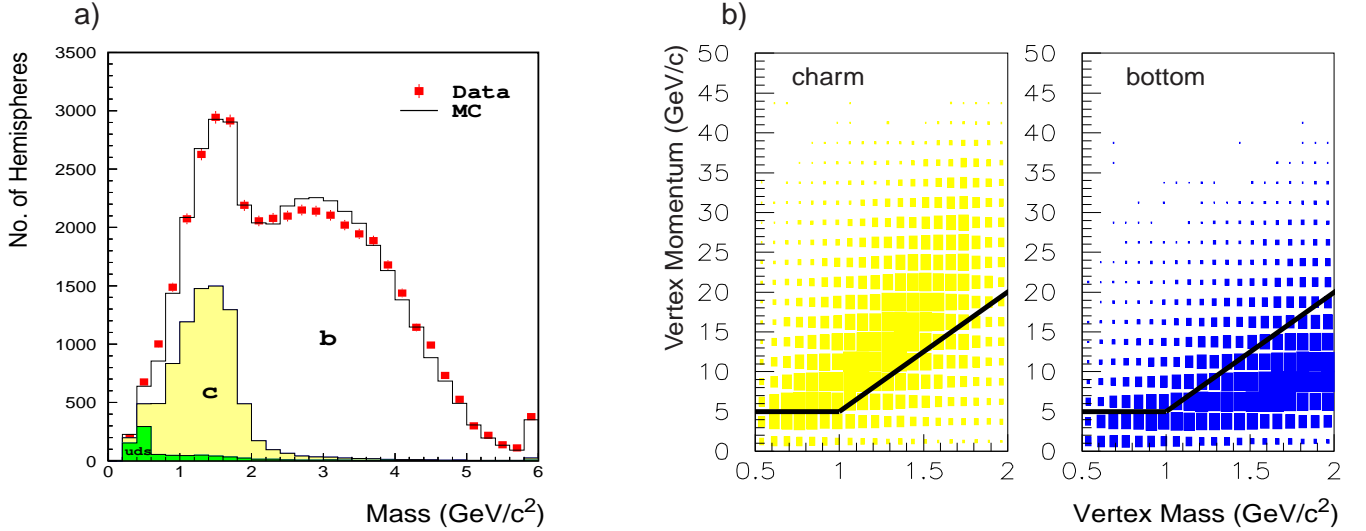} }   
\vskip 10pt
\caption{
a) Distributions of the $P_T$ corrected vertex mass for data (points)
and Monte Carlo prediction of b, c and uds. 
b) Scatter plots of vertex momentum and
mass for c (left) and b (right) events.  }
\end{figure}

Charm tag relies on the intermediate mass region 
($0.55\ GeV/c^2<M_{P_T}<2\ GeV/c^2$).
Additional separation is provided by the 2-dimensional cut in the
momentum-mass plane for the secondary vertex,
as shown in FIG.~1-b) The cuts of $P_{vtx}>5\ GeV/c$ and 
$15M_{vtx}-P_{vtx}<10$  provides 
70\% purity and 16\% efficiency for $Z^0\rightarrow c\bar{c}$ events. 

\section{Measurements of $R_b$ and $R_c$}

The SLD $R_b$ measurement is based on the double-tag technique\cite{rbprl}.  
Events are divided into two hemispheres 
by the plane perpendicular to the thrust axis of the event,
and a $b$-tag algorithm is applied to each hemisphere in turn. 
The fraction of hemispheres tagged as originating from $b$-quarks 
(single-tag) is given by
$$F_s = R_b\epsilon_b + R_c\epsilon_c + (1 - R_c - R_b)\epsilon_{uds}\ ,$$
and the fraction of events with both hemispheres tagged as originating
from $b$-quarks (double-tag) is given by
$$F_d = R_b({\epsilon_b}^2 + \lambda_b(\epsilon_b-\epsilon_b^2)) + 
R_c({\epsilon_c}^2 + \lambda_c(\epsilon_c-\epsilon_c^2)) + 
(1 - R_c - R_b){\epsilon_{uds}}^2.$$
The above two equations are solved for both $R_b$ and the $b$-tag efficiency
$\epsilon_b$.
The background tagging efficiencies for $uds$- and $c$-hemispheres, 
$\epsilon_{uds}$ and $\epsilon_c$, as well as the $b$-tag hemisphere correlation 
$\lambda_b=(\epsilon^{double}_b-\epsilon^2_b)/(\epsilon_b-\epsilon^2_b)$ 
are estimated from the Monte-Carlo. 
$R_c$ is assumed to be a Standard Model value.

For the $R_c$ measurement, the double-tag technique is extended 
to include both charm and bottom tags\cite{SLDRc}. 
Using the similar equations as above, we add the fraction of hemispheres 
tagged as originating from $c$-quarks
$$G_s = R_b\eta_b + R_c\eta_c + (1 - R_c - R_b)\eta_{uds}\ ,$$
and the fraction of events with both hemispheres tagged as originating
from $c$-quarks
$$G_d = R_b({\eta_b}^2 + \lambda^{\prime}_b(\eta_b-\eta_b^2)) + 
R_c({\eta_c}^2 + \lambda^{\prime}_c(\eta_c-\eta_c^2)) + 
(1 - R_c - R_b){\eta_{uds}}^2.$$
In the $R_c$ measurement, we have one more fraction of events where
one hemisphere is tagged as $b$ and another hemisphere is tagged as $c$
(mixed-tag)  
$$M = 2 \left[ R_b\epsilon_b\eta_b + R_c\epsilon_c\eta_c
      + (1 - R_c - R_b)\epsilon_{uds}\eta_{uds}\right].$$
The last three equations are solved for $R_c$, $c$-tag efficiency
$\eta_c$, and $b$-tag efficiency $\eta_b$.
Where the $uds$ efficiency $\eta_{uds}$ and the
correlations are taken from the Monte Carlo. $R_b$ and $\epsilon_b$
are known from the first two equations.
In general, a high purity tag is needed for a double-tag measurement.
However, the residual background in the $c$-tag sample are mainly $b$'s and
the mixed-tag equation allows us to solve $\eta_b$ from the data,
using the high purity $b$-tag in the opposite hemisphere.

\begin{figure}[ht]
\parbox{250pt}{
\vskip 15pt
  \parbox{225pt}{
    \epsfxsize 225pt
    \epsfysize 3.3in
    \epsfbox{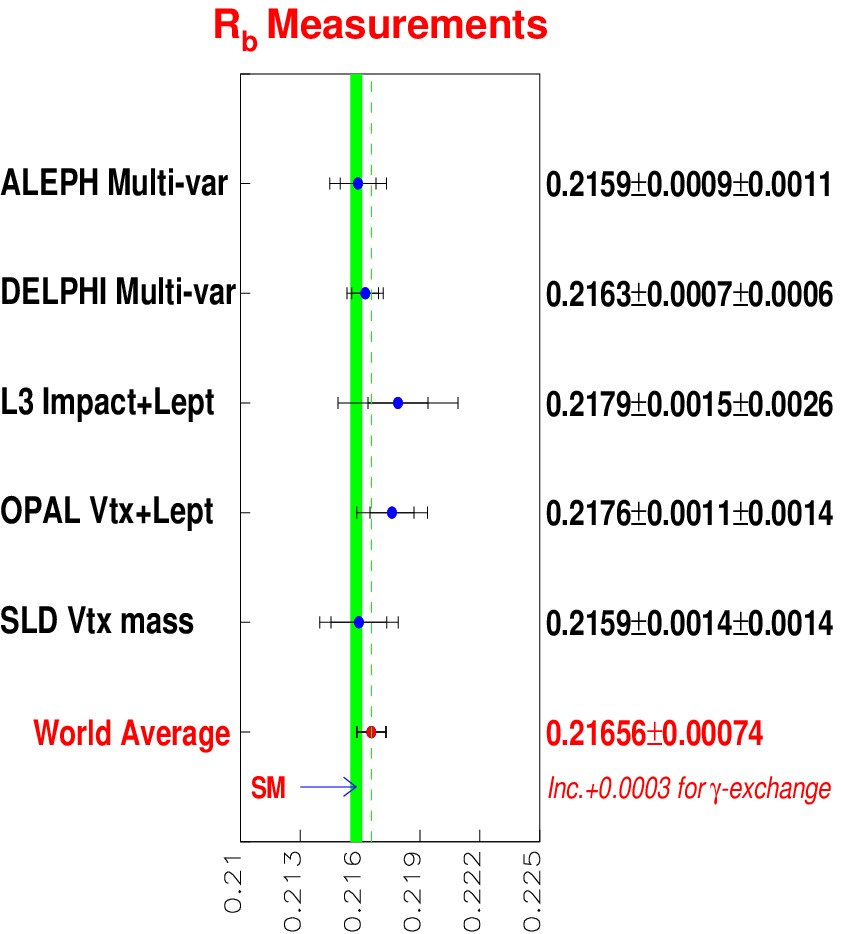}
  }
} 
\parbox{245pt}{
\vskip 15pt
  \parbox{220pt}{
    \epsfxsize 220pt
    \epsfysize 3.3in
    \epsfbox{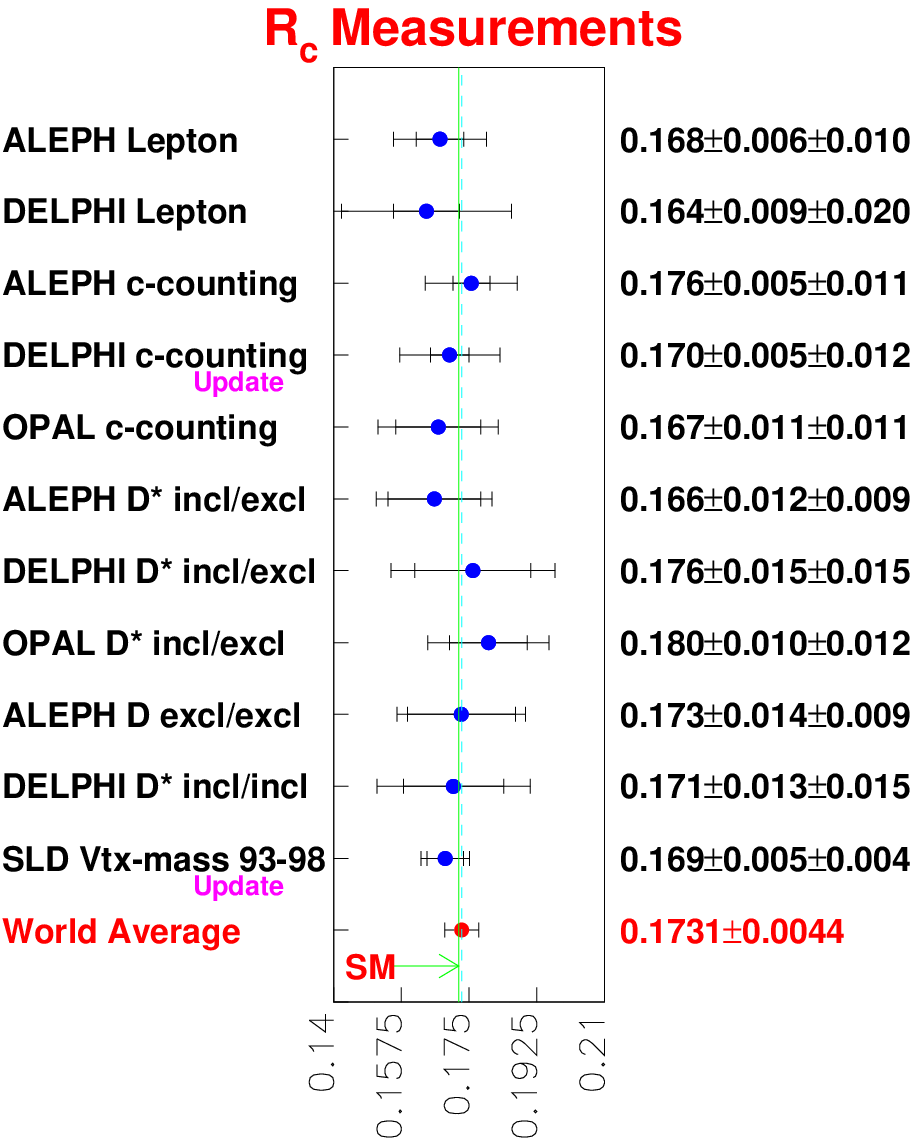}
  }
} 
\vskip 18pt
\caption{
Comparison of world $R_b$ (left) and $R_c$ (right)
    measurements. The inner and outer error bars present statistical and
    total errors, respectively.
}
\end{figure}

The SLD preliminary results of 
\begin{eqnarray}
R_b &=& 0.2159\pm 0.0014 (\mbox{stat.})\pm 0.0014 (\mbox{syst.})
 \nonumber\\
R_c &=& 0.1685 \pm 0.0047 (\mbox{stat.})\pm 0.0043 (\mbox{syst.}),
\nonumber
\end{eqnarray}
are obtained from the 1993-98 winter SLD run (400k $Z^0$) and the 1993-98 whole
run (550k $Z^0$), respectively.
Both results are in good agreement with the Standard-Model predictions. 
The largest uncertainties in $R_b$ and $R_c$ measurements are 
detector systematics, and Monte-Carlo statistics of the uds background, 
respectively.
FIG.~2. shows the comparison of the preliminary results 
of $R_b$ and $R_c$ measurements from the SLD and LEP experiments.

\section{$A_c$ measurements}
$A_f$ can be extracted by forming the forward-backward asymmetry
$$ A_{FB}^f(z) = \frac{\sigma^f(z)-\sigma^f(-z)}{\sigma^f(z)+\sigma^f(-z)}
            = A_e A_f \frac{2z}{1+z^2},$$
where $z = \cos\theta$ is the 
direction of the outgoing fermion relative to the incident electron.
$A_{FB}$ for quarks depends on both the initial state 
$A_e$ and the final state $A_f$. 
At the SLC, the ability to manipulate the longitudinal polarization 
of the electron beam allows the isolation
of the parameter $A_f$ independently of the $A_e$, 
through formation of the left-right forward-backward
double asymmetry:
$$ \tilde{A}_{FB}^f(z) = 
 \frac{[\sigma^f_L(z)-\sigma^f_L(-z)]-[\sigma^f_R(z)-\sigma^f_R(-z)]}
      {[\sigma^f_L(z)+\sigma^f_L(-z)]+[\sigma^f_R(z)+\sigma^f_R(-z)]}
               = |P_e| A_f \frac{2z}{1+z^2},$$
where $P_e$ is the longitudinal polarization of the electron beam. 
The high polarization of $\sim$77\% at the SLC also provides a large 
statistical advantage of $(P_e/A_e)^2\sim25$ compared to the $A_{FB}^f$ 
on the sensitivity to $A_f$.

In the actual analyses, we use an unbinned maximum likelihood fit based on
the Born-level cross section for fermion production in $Z^0$-boson
decay, to extract the $A_c$, instead of using the double asymmetry.
The likelihood function used in the analyses is
\begin{eqnarray}
\ln{\cal L}= \sum^{n}_{i=1} 
& \ln & \{
f_c \cdot [(1-P_eA_e)(1+z_i^2)+2(A_e-P_e)z_i \cdot A_c 
\cdot (1-\Delta_{QCD}^c(z_i))] \nonumber \\
 & + &  
f_b \cdot [(1-P_eA_e)(1+z_i^2)+2(A_e-P_e)z_i\cdot A_b 
\cdot (1-2\bar{\chi})
\cdot (1-\Delta_{QCD}^b(z_i))
] \nonumber \\
 & + &  
f_{BG} \cdot [(1+z_i^2)+2A_{BG}z_i]
\}
\nonumber 
\end{eqnarray}
where 
$n$ is the total number of candidates, 
$f_c$, $f_b$,  and $f_{BG}$ indicates the probabilities 
that a candidate is a signal from $c\bar{c}$, $b\bar{b}$,
or background, respectively. 
$\bar{\chi}$ is the $B^0\bar{B^0}$ mixing parameter, and 
$\Delta_{QCD}^f(y)$ is the $O(\alpha_s)$ QCD correction to the
asymmetry. 

At the SLD, four different techniques are used to measure the $A_c$:
inclusive charm-asymmetry measurement with Kaon charge and Vertex
charge, lepton, exclusively reconstructed D* and D-mesons, and a
new method using inclusive soft-pion from D*.
An inclusive charm tag using intermediate
vertex mass is used to select charm events in a similar manner as the 
SLD $R_c$ analysis\cite{SLDAcinc}. A $b$ veto is also applied to reject any event with
high vertex mass in either hemisphere. For the
hemispheres with a secondary vertex, a secondary track identified as 
$K^\pm$ from the CRID, or a non-zero vertex charge, is used to sign the 
charm quark direction.
The background is mostly $b$ events and its fraction is constrained by 
the double-tag calibration as for the $R_c$ measurement.
The preliminary result from the 1993-98
data sample is $A_c=0.603 \pm0.028(\mbox{stat.}) \pm0.023 (\mbox{syst.})$. 
This analysis has significantly high statistical power 
and the systematic errors are still very much under control. 

We also measure the charm asymmetry with traditional technique 
using electrons and muons which not only tag the $c$ events 
but also determine the $c$-quark direction from the lepton\cite{SLDAclepton}.
We get the preliminary result 
$A_c=0.567 \pm 0.051 (\mbox{stat.}) \pm 0.064 (\mbox{syst.})$ 
from the 1993-98 (muon) and 1993-97 (electron) SLD data.

The exclusive reconstruction of charmed mesons provide the cleanest
technique for the charm-asymmetry measurements\cite{SLDAcexcl}.
We use four decay modes to identify $D^{\ast+}$:
the decay $D^{\ast+} \rightarrow \pi_s^+ D^0 $ followed by 
$D^0 \rightarrow K^- \pi^+$,       
$D^0 \rightarrow K^- \pi^+ \pi^0$  (Satellite resonance),
$D^0 \rightarrow K^- \pi^+ \pi^- \pi^+$, or
$D^0 \rightarrow K^- l^+ \nu_l$ ($l=$e or $\mu$).
We also identify $D^+$ and $D^0$ mesons via the decay of
$D^+ \rightarrow K^- \pi^+ \pi^+$ and        
$D^0 \rightarrow K^- \pi^+$ (not from $D^{\ast+}$).       
In this analysis, we reject $Z^0 \rightarrow b\bar{b}$
events using $P_T$-corrected mass of the reconstructed vertices.
We required that reconstructed vertices had
a mass of less than 2.0 GeV/c$^2$. 
This cut rejected 57\% of 
$b\bar{b}$ events with 99\% of the remaining being $c\bar{c}$ events.
The random-combinatoric background can be estimated 
from the mass sidebands. 
The SLD preliminary result from this analysis using 550k of data 
from 1993-98 runs is 
$A_c=0.690 \pm 0.042 (\mbox{stat.}) \pm 0.022 (\mbox{syst.})$.

A new analysis using inclusive soft-pion from D* has been introduced 
by SLD in Winter-99. 
Since the decay $D^{\ast+}\rightarrow D^0 \pi_s$ has small
Q value of $m_{D^\ast} - m_{D^0} - m_{\pi}$ = 6 MeV$/c^2$,
the maximum transverse momentum of $\pi_s$ with respect to the 
$D^{\ast}$ flight direction is only 40 MeV.
To determine the $D^{\ast}$ direction, charged tracks and neutral
clusters are clustered into jets, using an invariant-mass algorithm.
We also reject the $b\bar{b}$ background using $P_T$-corrected-mass
information of reconstructed vertices.
The background shape was determined by the function of 
$F_{BG}(P_T^2) = a / (1 + bP_T^2 + c(P_T^2)^2)$.
FIG.~3. shows the $P_T^2$ distribution for the soft-pion
tracks. 
The region of $P_T^2 < 0.01$ (GeV/c)$^2$ is regarded as a
signal region, where a signal-to-background ratio of 1:2 is observed.
From the 1993-98 SLD data, we get the preliminary result of 
$A_c=0.683 \pm 0.052 (\mbox{stat.}) \pm 0.050 (\mbox{syst.})$.
The largest systematic uncertainty is the choice of background shape.

\parbox{245pt}{
\vskip 15pt
\vskip 0.3in
  \parbox{180pt}{
    \epsfxsize 180pt
    \epsfysize 2.6in
    \epsfbox{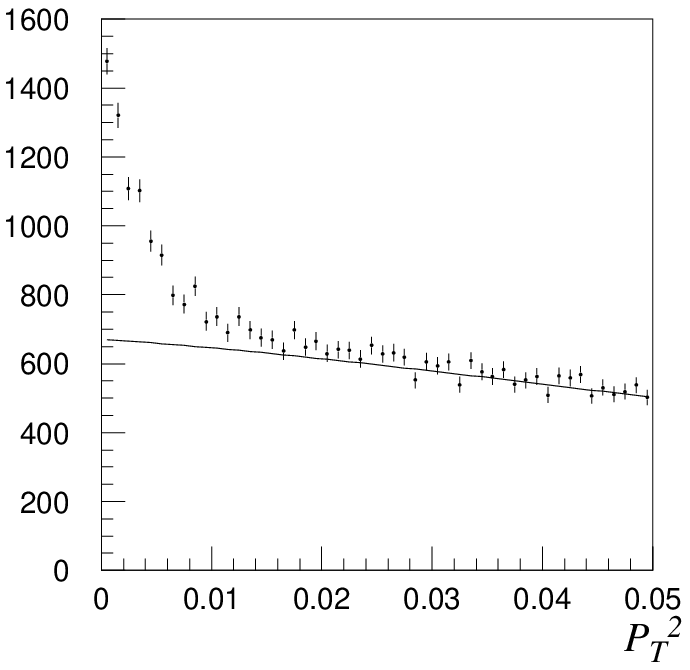}
  }
  \vskip 0.4in
  \parbox{220pt}{
    {\footnotesize    \hspace{10pt}
    FIG.~3. $P_T^2$ distribution for 
    the soft-pion tracks. Background shape is
    obtained by the function described in the text. }
  }
} 
\parbox{245pt}{
\vskip 15pt
  \parbox{220pt}{
    \epsfxsize 220pt
    \epsfysize 3.3in
    \epsfbox{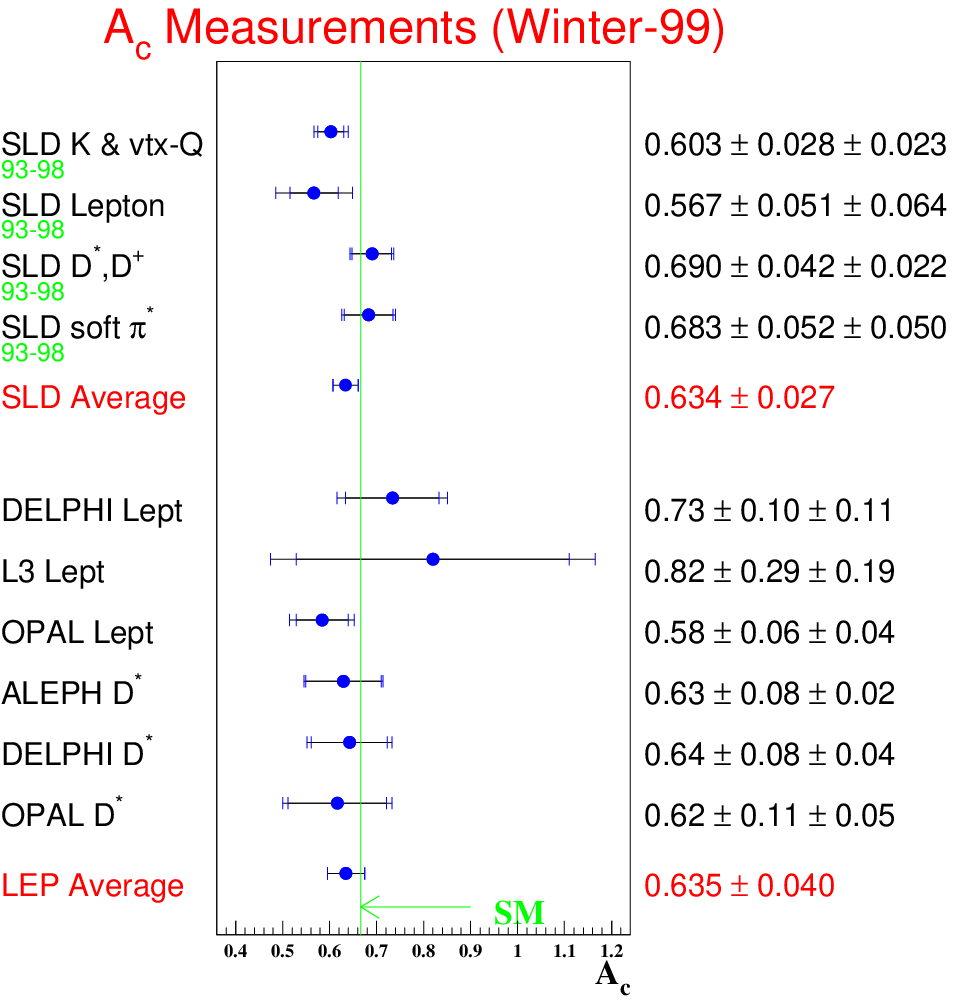}
  }
  \parbox{220pt}{
  {\footnotesize \hspace{10pt}
   FIG.~4. Comparison of world $A_c$ measurements. 
   The inner and outer error bars present statistical and
   total errors, respectively.}
  }
}

\vskip 20pt
FIG.~4. shows the preliminary results from the SLD and LEP
measurements, where the LEP
measurements are derived from $A_c = 4A_{FB}^{0,b}$/$\left( 3A_e
\right)$ using $A_e = 0.1491 \pm 0.0018$ (the combined SLD $A_{LR}$
and LEP $A_{lepton}$). 
The combined preliminary SLD result for $A_c$ is obtained as 
$$A_c = 0.634 \pm 0.027 .$$

\section{CONCLUSION}

SLD produces world class electroweak-parameter measurements 
in the heavy-quark sector.
The SLD measurements of $R_c$ and $A_c$ are now the most 
precise single measurements in the world. 
The measured $R_b$, $R_c$ and $A_c$ results are consistent with the 
Standard Model, and some analyses will improve when
the full set of 1993-98 SLD data is included.




\section*{$^{**}$List of Authors} 


%
%
%
\begin{center}
\def\iADEL{$^{(1)}$}
\def\iAOMORI{$^{(2)}$}
\def\iBOLO{$^{(3)}$}
\def\iBRI{$^{(4)}$}
\def\iBRUN{$^{(5)}$}
\def\iBU{$^{(6)}$}
\def\iCINC{$^{(7)}$}
\def\iCOLO{$^{(8)}$}
\def\iCOLU{$^{(9)}$}
\def\iCSU{$^{(10)}$}
\def\iFERR{$^{(11)}$}
\def\iFRAS{$^{(12)}$}
\def\iILLI{$^{(13)}$}
\def\iJHU{$^{(14)}$}
\def\iLBL{$^{(15)}$}
\def\iLTU{$^{(16)}$}
\def\iMASS{$^{(17)}$}
\def\iMISSI{$^{(18)}$}
\def\iMIT{$^{(19)}$}
\def\iMOSCOW{$^{(20)}$}
\def\iNAGO{$^{(21)}$}
\def\iOREG{$^{(22)}$}
\def\iOXF{$^{(23)}$}
\def\iPADO{$^{(24)}$}
\def\iPERU{$^{(25)}$}
\def\iPISA{$^{(26)}$}
\def\iRAL{$^{(27)}$}
\def\iRUTG{$^{(28)}$}
\def\iSLAC{$^{(29)}$}
\def\iSOGA{$^{(30)}$}
\def\iSOONG{$^{(31)}$}
\def\iTENN{$^{(32)}$}
\def\iTOHO{$^{(33)}$}
\def\iUCSB{$^{(34)}$}
\def\iUCSC{$^{(35)}$}
\def\iUVIC{$^{(36)}$}
\def\iVAND{$^{(37)}$}
\def\iWASH{$^{(38)}$}
\def\iWISC{$^{(39)}$}
\def\iYALE{$^{(40)}$}

  \baselineskip=.75\baselineskip  
\mbox{Kenji  Abe\unskip,\iNAGO}
\mbox{Koya Abe\unskip,\iTOHO}
\mbox{T. Abe\unskip,\iSLAC}
\mbox{I. Adam\unskip,\iSLAC}
\mbox{T.  Akagi\unskip,\iSLAC}
\mbox{N.J. Allen\unskip,\iBRUN}
\mbox{W.W. Ash\unskip,\iSLAC}
\mbox{D. Aston\unskip,\iSLAC}
\mbox{K.G. Baird\unskip,\iMASS}
\mbox{C. Baltay\unskip,\iYALE}
\mbox{H.R. Band\unskip,\iWISC}
\mbox{M.B. Barakat\unskip,\iLTU}
\mbox{O. Bardon\unskip,\iMIT}
\mbox{T.L. Barklow\unskip,\iSLAC}
\mbox{G.L. Bashindzhagyan\unskip,\iMOSCOW}
\mbox{J.M. Bauer\unskip,\iMISSI}
\mbox{G. Bellodi\unskip,\iOXF}
\mbox{R. Ben-David\unskip,\iYALE}
\mbox{A.C. Benvenuti\unskip,\iBOLO}
\mbox{G.M. Bilei\unskip,\iPERU}
\mbox{D. Bisello\unskip,\iPADO}
\mbox{G. Blaylock\unskip,\iMASS}
\mbox{J.R. Bogart\unskip,\iSLAC}
\mbox{G.R. Bower\unskip,\iSLAC}
\mbox{J.E. Brau\unskip,\iOREG}
\mbox{M. Breidenbach\unskip,\iSLAC}
\mbox{W.M. Bugg\unskip,\iTENN}
\mbox{D. Burke\unskip,\iSLAC}
\mbox{T.H. Burnett\unskip,\iWASH}
\mbox{P.N. Burrows\unskip,\iOXF}
\mbox{A. Calcaterra\unskip,\iFRAS}
\mbox{D. Calloway\unskip,\iSLAC}
\mbox{B. Camanzi\unskip,\iFERR}
\mbox{M. Carpinelli\unskip,\iPISA}
\mbox{R. Cassell\unskip,\iSLAC}
\mbox{R. Castaldi\unskip,\iPISA}
\mbox{A. Castro\unskip,\iPADO}
\mbox{M. Cavalli-Sforza\unskip,\iUCSC}
\mbox{A. Chou\unskip,\iSLAC}
\mbox{E. Church\unskip,\iWASH}
\mbox{H.O. Cohn\unskip,\iTENN}
\mbox{J.A. Coller\unskip,\iBU}
\mbox{M.R. Convery\unskip,\iSLAC}
\mbox{V. Cook\unskip,\iWASH}
\mbox{R.F. Cowan\unskip,\iMIT}
\mbox{D.G. Coyne\unskip,\iUCSC}
\mbox{G. Crawford\unskip,\iSLAC}
\mbox{C.J.S. Damerell\unskip,\iRAL}
\mbox{M.N. Danielson\unskip,\iCOLO}
\mbox{M. Daoudi\unskip,\iSLAC}
\mbox{N. de Groot\unskip,\iBRI}
\mbox{R. Dell'Orso\unskip,\iPERU}
\mbox{P.J. Dervan\unskip,\iBRUN}
\mbox{R. de Sangro\unskip,\iFRAS}
\mbox{M. Dima\unskip,\iCSU}
\mbox{A. D'Oliveira\unskip,\iCINC}
\mbox{D.N. Dong\unskip,\iMIT}
\mbox{M. Doser\unskip,\iSLAC}
\mbox{R. Dubois\unskip,\iSLAC}
\mbox{B.I. Eisenstein\unskip,\iILLI}
\mbox{V. Eschenburg\unskip,\iMISSI}
\mbox{E. Etzion\unskip,\iWISC}
\mbox{S. Fahey\unskip,\iCOLO}
\mbox{D. Falciai\unskip,\iFRAS}
\mbox{C. Fan\unskip,\iCOLO}
\mbox{J.P. Fernandez\unskip,\iUCSC}
\mbox{M.J. Fero\unskip,\iMIT}
\mbox{K. Flood\unskip,\iMASS}
\mbox{R. Frey\unskip,\iOREG}
\mbox{J. Gifford\unskip,\iUVIC}
\mbox{T. Gillman\unskip,\iRAL}
\mbox{G. Gladding\unskip,\iILLI}
\mbox{S. Gonzalez\unskip,\iMIT}
\mbox{E.R. Goodman\unskip,\iCOLO}
\mbox{E.L. Hart\unskip,\iTENN}
\mbox{J.L. Harton\unskip,\iCSU}
\mbox{A. Hasan\unskip,\iBRUN}
\mbox{K. Hasuko\unskip,\iTOHO}
\mbox{S.J. Hedges\unskip,\iBU}
\mbox{S.S. Hertzbach\unskip,\iMASS}
\mbox{M.D. Hildreth\unskip,\iSLAC}
\mbox{J. Huber\unskip,\iOREG}
\mbox{M.E. Huffer\unskip,\iSLAC}
\mbox{E.W. Hughes\unskip,\iSLAC}
\mbox{X. Huynh\unskip,\iSLAC}
\mbox{H. Hwang\unskip,\iOREG}
\mbox{M. Iwasaki\unskip,\iOREG}
\mbox{D.J. Jackson\unskip,\iRAL}
\mbox{P. Jacques\unskip,\iRUTG}
\mbox{J.A. Jaros\unskip,\iSLAC}
\mbox{Z.Y. Jiang\unskip,\iSLAC}
\mbox{A.S. Johnson\unskip,\iSLAC}
\mbox{J.R. Johnson\unskip,\iWISC}
\mbox{R.A. Johnson\unskip,\iCINC}
\mbox{T. Junk\unskip,\iSLAC}
\mbox{R. Kajikawa\unskip,\iNAGO}
\mbox{M. Kalelkar\unskip,\iRUTG}
\mbox{Y. Kamyshkov\unskip,\iTENN}
\mbox{H.J. Kang\unskip,\iRUTG}
\mbox{I. Karliner\unskip,\iILLI}
\mbox{H. Kawahara\unskip,\iSLAC}
\mbox{Y.D. Kim\unskip,\iSOGA}
\mbox{M.E. King\unskip,\iSLAC}
\mbox{R. King\unskip,\iSLAC}
\mbox{R.R. Kofler\unskip,\iMASS}
\mbox{N.M. Krishna\unskip,\iCOLO}
\mbox{R.S. Kroeger\unskip,\iMISSI}
\mbox{M. Langston\unskip,\iOREG}
\mbox{A. Lath\unskip,\iMIT}
\mbox{D.W.G. Leith\unskip,\iSLAC}
\mbox{V. Lia\unskip,\iMIT}
\mbox{C.Lin\unskip,\iMASS}
\mbox{M.X. Liu\unskip,\iYALE}
\mbox{X. Liu\unskip,\iUCSC}
\mbox{M. Loreti\unskip,\iPADO}
\mbox{A. Lu\unskip,\iUCSB}
\mbox{H.L. Lynch\unskip,\iSLAC}
\mbox{J. Ma\unskip,\iWASH}
\mbox{G. Mancinelli\unskip,\iRUTG}
\mbox{S. Manly\unskip,\iYALE}
\mbox{G. Mantovani\unskip,\iPERU}
\mbox{T.W. Markiewicz\unskip,\iSLAC}
\mbox{T. Maruyama\unskip,\iSLAC}
\mbox{H. Masuda\unskip,\iSLAC}
\mbox{E. Mazzucato\unskip,\iFERR}
\mbox{A.K. McKemey\unskip,\iBRUN}
\mbox{B.T. Meadows\unskip,\iCINC}
\mbox{G. Menegatti\unskip,\iFERR}
\mbox{R. Messner\unskip,\iSLAC}
\mbox{P.M. Mockett\unskip,\iWASH}
\mbox{K.C. Moffeit\unskip,\iSLAC}
\mbox{T.B. Moore\unskip,\iYALE}
\mbox{M.Morii\unskip,\iSLAC}
\mbox{D. Muller\unskip,\iSLAC}
\mbox{V. Murzin\unskip,\iMOSCOW}
\mbox{T. Nagamine\unskip,\iTOHO}
\mbox{S. Narita\unskip,\iTOHO}
\mbox{U. Nauenberg\unskip,\iCOLO}
\mbox{H. Neal\unskip,\iSLAC}
\mbox{M. Nussbaum\unskip,\iCINC}
\mbox{N. Oishi\unskip,\iNAGO}
\mbox{D. Onoprienko\unskip,\iTENN}
\mbox{L.S. Osborne\unskip,\iMIT}
\mbox{R.S. Panvini\unskip,\iVAND}
\mbox{C.H. Park\unskip,\iSOONG}
\mbox{T.J. Pavel\unskip,\iSLAC}
\mbox{I. Peruzzi\unskip,\iFRAS}
\mbox{M. Piccolo\unskip,\iFRAS}
\mbox{L. Piemontese\unskip,\iFERR}
\mbox{K.T. Pitts\unskip,\iOREG}
\mbox{R.J. Plano\unskip,\iRUTG}
\mbox{R. Prepost\unskip,\iWISC}
\mbox{C.Y. Prescott\unskip,\iSLAC}
\mbox{G.D. Punkar\unskip,\iSLAC}
\mbox{J. Quigley\unskip,\iMIT}
\mbox{B.N. Ratcliff\unskip,\iSLAC}
\mbox{T.W. Reeves\unskip,\iVAND}
\mbox{J. Reidy\unskip,\iMISSI}
\mbox{P.L. Reinertsen\unskip,\iUCSC}
\mbox{P.E. Rensing\unskip,\iSLAC}
\mbox{L.S. Rochester\unskip,\iSLAC}
\mbox{P.C. Rowson\unskip,\iCOLU}
\mbox{J.J. Russell\unskip,\iSLAC}
\mbox{O.H. Saxton\unskip,\iSLAC}
\mbox{T. Schalk\unskip,\iUCSC}
\mbox{R.H. Schindler\unskip,\iSLAC}
\mbox{B.A. Schumm\unskip,\iUCSC}
\mbox{J. Schwiening\unskip,\iSLAC}
\mbox{S. Sen\unskip,\iYALE}
\mbox{V.V. Serbo\unskip,\iSLAC}
\mbox{M.H. Shaevitz\unskip,\iCOLU}
\mbox{J.T. Shank\unskip,\iBU}
\mbox{G. Shapiro\unskip,\iLBL}
\mbox{D.J. Sherden\unskip,\iSLAC}
\mbox{K.D. Shmakov\unskip,\iTENN}
\mbox{C. Simopoulos\unskip,\iSLAC}
\mbox{N.B. Sinev\unskip,\iOREG}
\mbox{S.R. Smith\unskip,\iSLAC}
\mbox{M.B. Smy\unskip,\iCSU}
\mbox{J.A. Snyder\unskip,\iYALE}
\mbox{H. Staengle\unskip,\iCSU}
\mbox{A. Stahl\unskip,\iSLAC}
\mbox{P. Stamer\unskip,\iRUTG}
\mbox{H. Steiner\unskip,\iLBL}
\mbox{R. Steiner\unskip,\iADEL}
\mbox{M.G. Strauss\unskip,\iMASS}
\mbox{D. Su\unskip,\iSLAC}
\mbox{F. Suekane\unskip,\iTOHO}
\mbox{A. Sugiyama\unskip,\iNAGO}
\mbox{S. Suzuki\unskip,\iNAGO}
\mbox{M. Swartz\unskip,\iJHU}
\mbox{A. Szumilo\unskip,\iWASH}
\mbox{T. Takahashi\unskip,\iSLAC}
\mbox{F.E. Taylor\unskip,\iMIT}
\mbox{J. Thom\unskip,\iSLAC}
\mbox{E. Torrence\unskip,\iMIT}
\mbox{N.K. Toumbas\unskip,\iSLAC}
\mbox{T. Usher\unskip,\iSLAC}
\mbox{C. Vannini\unskip,\iPISA}
\mbox{J. Va'vra\unskip,\iSLAC}
\mbox{E. Vella\unskip,\iSLAC}
\mbox{J.P. Venuti\unskip,\iVAND}
\mbox{R. Verdier\unskip,\iMIT}
\mbox{P.G. Verdini\unskip,\iPISA}
\mbox{D.L. Wagner\unskip,\iCOLO}
\mbox{S.R. Wagner\unskip,\iSLAC}
\mbox{A.P. Waite\unskip,\iSLAC}
\mbox{S. Walston\unskip,\iOREG}
\mbox{J. Wang\unskip,\iSLAC}
\mbox{S.J. Watts\unskip,\iBRUN}
\mbox{A.W. Weidemann\unskip,\iTENN}
\mbox{E. R. Weiss\unskip,\iWASH}
\mbox{J.S. Whitaker\unskip,\iBU}
\mbox{S.L. White\unskip,\iTENN}
\mbox{F.J. Wickens\unskip,\iRAL}
\mbox{B. Williams\unskip,\iCOLO}
\mbox{D.C. Williams\unskip,\iMIT}
\mbox{S.H. Williams\unskip,\iSLAC}
\mbox{S. Willocq\unskip,\iMASS}
\mbox{R.J. Wilson\unskip,\iCSU}
\mbox{W.J. Wisniewski\unskip,\iSLAC}
\mbox{J. L. Wittlin\unskip,\iMASS}
\mbox{M. Woods\unskip,\iSLAC}
\mbox{G.B. Word\unskip,\iVAND}
\mbox{T.R. Wright\unskip,\iWISC}
\mbox{J. Wyss\unskip,\iPADO}
\mbox{R.K. Yamamoto\unskip,\iMIT}
\mbox{J.M. Yamartino\unskip,\iMIT}
\mbox{X. Yang\unskip,\iOREG}
\mbox{J. Yashima\unskip,\iTOHO}
\mbox{S.J. Yellin\unskip,\iUCSB}
\mbox{C.C. Young\unskip,\iSLAC}
\mbox{H. Yuta\unskip,\iAOMORI}
\mbox{G. Zapalac\unskip,\iWISC}
\mbox{R.W. Zdarko\unskip,\iSLAC}
\mbox{J. Zhou\unskip.\iOREG}

\it
  \vskip \baselineskip                   
  \centerline{(The SLD Collaboration)}   
  \vskip \baselineskip        
  \baselineskip=.75\baselineskip   
\iADEL
  Adelphi University, Garden City, New York 11530, \break
\iAOMORI
  Aomori University, Aomori , 030 Japan, \break
\iBOLO
  INFN Sezione di Bologna, I-40126, Bologna, Italy, \break
\iBRI
  University of Bristol, Bristol, U.K., \break
\iBRUN
  Brunel University, Uxbridge, Middlesex, UB8 3PH United Kingdom, \break
\iBU
  Boston University, Boston, Massachusetts 02215, \break
\iCINC
  University of Cincinnati, Cincinnati, Ohio 45221, \break
\iCOLO
  University of Colorado, Boulder, Colorado 80309, \break
\iCOLU
  Columbia University, New York, New York 10533, \break
\iCSU
  Colorado State University, Ft. Collins, Colorado 80523, \break
\iFERR
  INFN Sezione di Ferrara and Universita di Ferrara, I-44100 Ferrara, Italy, \break
\iFRAS
  INFN Lab. Nazionali di Frascati, I-00044 Frascati, Italy, \break
\iILLI
  University of Illinois, Urbana, Illinois 61801, \break
\iJHU
  Johns Hopkins University,  Baltimore, Maryland 21218-2686, \break
\iLBL
  Lawrence Berkeley Laboratory, University of California, Berkeley, California 94720, \break
\iLTU
  Louisiana Technical University, Ruston,Louisiana 71272, \break
\iMASS
  University of Massachusetts, Amherst, Massachusetts 01003, \break
\iMISSI
  University of Mississippi, University, Mississippi 38677, \break
\iMIT
  Massachusetts Institute of Technology, Cambridge, Massachusetts 02139, \break
\iMOSCOW
  Institute of Nuclear Physics, Moscow State University, 119899, Moscow Russia, \break
\iNAGO
  Nagoya University, Chikusa-ku, Nagoya, 464 Japan, \break
\iOREG
  University of Oregon, Eugene, Oregon 97403, \break
\iOXF
  Oxford University, Oxford, OX1 3RH, United Kingdom, \break
\iPADO
  INFN Sezione di Padova and Universita di Padova I-35100, Padova, Italy, \break
\iPERU
  INFN Sezione di Perugia and Universita di Perugia, I-06100 Perugia, Italy, \break
\iPISA
  INFN Sezione di Pisa and Universita di Pisa, I-56010 Pisa, Italy, \break
\iRAL
  Rutherford Appleton Laboratory, Chilton, Didcot, Oxon OX11 0QX United Kingdom, \break
\iRUTG
  Rutgers University, Piscataway, New Jersey 08855, \break
\iSLAC
  Stanford Linear Accelerator Center, Stanford University, Stanford, California 94309, \break
\iSOGA
  Sogang University, Seoul, Korea, \break
\iSOONG
  Soongsil University, Seoul, Korea 156-743, \break
\iTENN
  University of Tennessee, Knoxville, Tennessee 37996, \break
\iTOHO
  Tohoku University, Sendai 980, Japan, \break
\iUCSB
  University of California at Santa Barbara, Santa Barbara, California 93106, \break
\iUCSC
  University of California at Santa Cruz, Santa Cruz, California 95064, \break
\iUVIC
  University of Victoria, Victoria, British Columbia, Canada V8W 3P6, \break
\iVAND
  Vanderbilt University, Nashville,Tennessee 37235, \break
\iWASH
  University of Washington, Seattle, Washington 98105, \break
\iWISC
  University of Wisconsin, Madison,Wisconsin 53706, \break
\iYALE
  Yale University, New Haven, Connecticut 06511. \break

\rm
%

\end{center}

\end{document}